\documentstyle[sprocl,epsf]{article}

\def\be{\begin{equation}}
\def\ee{\end{equation}}
\def\bea{\begin{eqnarray}}
\def\eea{\end{eqnarray}}
\begin{document}
\title{CRITICAL PHENOMENA IN FINITE SYSTEMS}
\author{A. BONASERA$^1$, T. MARUYAMA$^{1,2}$ and S. CHIBA$^2$}
\address{$^1$Laboratorio Nazionale del Sud, Istituto Nazionale di Fisica
Nucleare,
           V.S.Sofia 44, 95123 Catania-Italy.\\
	   $^2$Advanced Science Research Center, Japan Atomic Energy Research Institute, Tokai,
	   Ibaraki, 319-1195, Japan.}
\maketitle\abstracts{ 
We discuss the dynamics of finite systems within molecular dynamics models.  Signatures of a 
critical behavior are analyzed and compared to experimental data both in nucleus-nucleus and
metallic cluster collisions.  We suggest the possibility to explore the instability region
via tunneling.  In this way we can obtain fragments at very low temperatures and densities.
We call these fragments quantum drops.}

\section{Introduction}
The dynamics of a fragmenting finite system can be well modeled by Classical
Molecular Dynamics (CMD).  Of course in such studies we are not interested
in reproducing the data coming from nucleus-nucleus, cluster-cluster or
fullerene-fullerene collisions, which are strongly influenced by quantum
features, but simply
in the possibility that a finite (un)charged system ``remembers'' of a 
liquid to gas phase transition which occurs in the infinite case 
limit~\cite{ref1}.  Classical particles
interacting through a short range repulsive interaction and a longer range
attractive one have an Equation of State (EOS) which resembles a Van Der
Waals (VDW)~\cite{ref1,ref2}.  It is also well known that the Nuclear EOS
 resembles a  VDW as well~\cite{ref3}, thus classical studies of the
instability region are quite justified in order to understand the finite size
plus Coulomb effects.

\begin{figure}[thb]
\epsfxsize=1.00\textwidth
\epsfbox{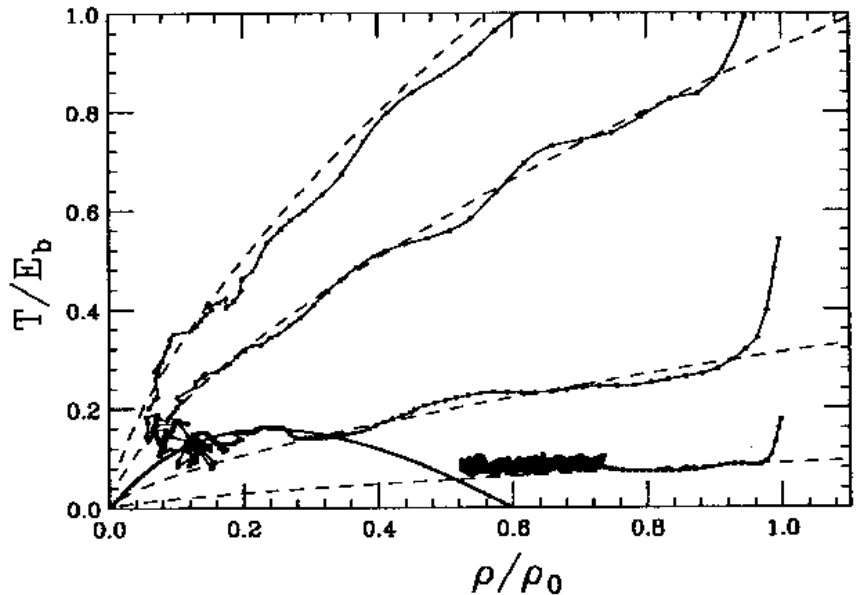}
\caption{Time evolution of the biggest fragment in the density temperature 
plane.}
\end{figure}
Starting from a two body force, for instance of the Yukawa type, the EOS can
be easily calculated within CMD using the virial theorem.  This was done
in ref.\cite {ref1} and the resulting EOS spinodal line is plotted in 
Fig.~1 (full line) in the temperature $T$ density $\rho$ plane. These quantities
 have
been normalized using typical values of the system.  When dealing with phase
transitions it is usual to normalize the various quantities by their values
at the critical point of a second order phase transitions.  Since later
on, we will compare with some results on finite systems,
where the values at the critical point (if it exist) are not known, we normalize
by typical values in the ground state of the system such as density $\rho_0$,
or the absolute value of the binding energy $E_b$.  In such a way we can 
compare results coming from different systems such as nuclei 
or clusters~\cite{schu}. 

Within CMD the time evolution of a finite and excited systems can be
 numerically calculated.  In ref.\cite{ref1} the time evolution of 100
particles was followed.  The particles were initially given a Maxewell
Boltzmann distribution at temperature $T$.  The density and temperature estimated
for the biggest fragment are plotted in Fig.~1 (full lines).  Each dot is
printed at regular time intervals of 1 fm/c.  We have used a gs density for
the finite system of $0.12 {\rm fm}^{-3}$ ($0.15 {\rm fm}^{-3}$ 
for the infinite case limit)
and a B.E. of $-10$ MeV ($-16$ MeV in the infinite case) 
to normalize the curves.
 Without such normalizations the lowest curve ($T=2$ MeV) for instance will enter
deeply inside the instability region which is not the case since this is a
typical case of evaporation.  
 ``Critical'' events
occur for the initial $T=5$ MeV~\cite{ref1}, which we see that enters into the
instability region near the critical point for a second order phase transition.
Note that the value of the excitation energy $E^*$ for this case is 
$E^*/E_b=0.71$, which coincides with the value of the $E^*/E_b$ obtained in
the infinite case limit at the critical density and temperature.  Some
features in this figure are of interest.  The first one is that the expansion
is isentropic (dashed lines in Fig.~1) until the system enters the 
instability region;  events at high $T$ do not enter the instability region, i.e.
the system dissolves quickly because of the very large excitation energy.
Already from these numerical experiments we understand that finite systems
are quite different from the infinite limit case, in that there is no
confining volume, thus $T$ and $\rho$ are time dependent quantities.  Their
values cannot be fixed from the outside but, to some extent the initial
excitation energy and density can be fixed. The concept of temperature
becomes questionable as well especially at high $E^*$.  In fact the
system expands rather quickly, and the very energetic particles leave
the system without interacting with other particles.  Thus fluctuations
are small and decrease as we will show later with increasing $E^*$. 
Looking more carefully at Fig.~1, we notice that it is quite difficult to explore
the instability region at small $T$.  In fact in the case $T=2$ MeV the system is not
 able to enter the instability region.  We need to give more excitation energy to it, but
 if we do so the $T$ when the system enters the instability region is larger.  We can 
 imagine to have a barrier in a collective space where the radial coordinate R and its
  conjugate momentum are the relevant degrees of freedom. Analogously to the process of
  spontaneous fission (SF) we can imagine that in presence of the long range Coulomb force, the
  barrier as a finite width, thus the quantum mechanical process of tunneling might be
  possible and the system enters the spinodal region at low $T$.  We call the fragments thus 
  formed quantum drops and discuss the process in a later section.
\section{Observables}

During a heavy ion collision many fragments are formed and finally detected.
The first observable that was tested was the mass
yield~\cite{eos}.
In fact from the
Fisher model of phase transition we expect that if there is a phase transition,
a power law should appear in the spectrum, see ref.\cite{fis}.
In order to illustrate the Fisher model and to have a look at the
modifications needed when dealing with a finite system we have repeated the
calculations described above, each time changing randomly the momenta of the 
particles.
%\begin{figure}
%\vspace{5cm}  % amount of vertical space needed
%\caption{Mass yields at different initial temperatures.}
%\end{figure}
In the simulation we can generate
thousands of events and produce mass distributions at each $T$.  
The mass distribution so obtained 
is in good agreement with the Fisher law.
The fits are rather good
especially near and above the $T=5$ MeV case.  The latter gives a very
good power law distribution with $\tau=2.23$.  Such a value of $\tau$ is
in good agreement with what observed in liquid-gas phase transitions of
finite systems.  At very high $T$ the yield is practically exponentially
decreasing, while low $T$ give one big fragment accompanied of some
 small ones.  
Other variables like moments of the mass distributions, 
Campi plots~\cite{cam86}, give
indications for a possible second order phase transition.
In ref.\cite{mast} we have tested if these findings based on CMD results have
some resemblance with reality.  The experiment Au+Au at 35 MeV/A performed at
MSU using the Multics-Miniball detectors~\cite{mast} has been analyzed in terms
of moments of mass distributions.  First the collision was simulated in the
framework of CMD and the effects of the experimental device were studied in
detail~\cite{belk}.  It was found that a good reconstruction of the PLF can
be obtained with this device.  Thus peripheral collisions are rather well
detected by the Multics-Miniball apparatus.  For such collisions a critical
behavior had been predicted within CMD in ref.\cite{dzp} and confirmed
for this system and at this energy as well~\cite{belk,rnc}. 

\section{Chaotic Dynamics} 

The observation of large fluctuations in fragmentation hints to the occurrence
of chaos. In order to
address this problem quantitatively we calculate the maximal Lyapunov Exponents
(MLE)
as a function of the initial excitation energy.
An important property of chaotic motion is the high sensibility to changes
in the initial conditions.
Closely neighboring trajectories diverge exponentially in time. For regular
trajectories, on the other hand , they are found to diverge only
linearly. The quantity that properly quantifies the rate of exponential
divergence are the LE~\cite{pp2}.  The MLE have been calculated 
in ref.\cite{pp2} for the system of Fig.~1 
and analogous calculations have been performed in
ref.\cite{burgio} within the Boltzmann Nordheim Vlasov (BNV) framework, i.e. a
mean field description of a disassembling nucleus. In Fig.~2 the LE is plotted
vs. excitation energy in the CMD (diamonds) and BNV case (squares).
The qualitative behavior is the same, i.e. both calculations display a
maximum at the normalized $E^*=0.5$.  
\begin{figure}[thb]
\epsfxsize=1.00\textwidth
\epsfbox{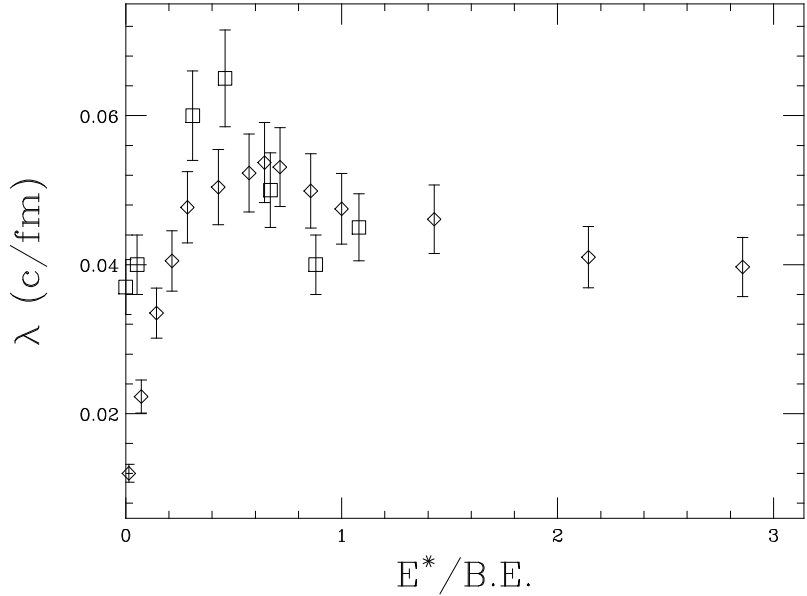}
\caption{Maximal Lyapunov exponents in CMD (diamonds) and BNV (squares) calculations.}
\end{figure}
Such maximum corresponds in the CMD 
to the value where the mass distribution is a power law.  At low excitation
energy the LE calculated in BNV are larger than the CMD case because the gs of
the nucleus is a liquid while the classical gs is a solid.  Very important is
the decrease of the LE at high $E^*$.  This clearly demonstrates that the
degree of chaoticity, i.e. thermalization is not increasing and the initial
excitation energy is partially thermal but a large amount is in the
form of collective expansion.  This is consistent with the picture of
a limiting temperature that the nucleus can sustain~\cite{hagel}.  
In fact we can have
a properly thermalized system when the self consistent field is able to bind
the particles in some volume for some time.  This field acts as a confining
volume where the particles stay to boil.  But, when the excitation energy is
large, particles have enough kinetic energy to leave the system promptly.
This picture greatly clarifies the dynamics of fragmentation.  At low
excitation energy we can have a liquid at a temperature $T$ which evaporates
particles.  At high $E^*$ energetic fragments are quickly emitted and a
small liquid at a limiting $T$ remains.  Thus the transition is from liquid
to free particles thus somewhat different from a liquid to gas phase transition
that would occur if the system was confined in a box.

\section{Quantum Drops}

\begin{figure}[thb]
\epsfxsize=0.98\textwidth
\epsfbox{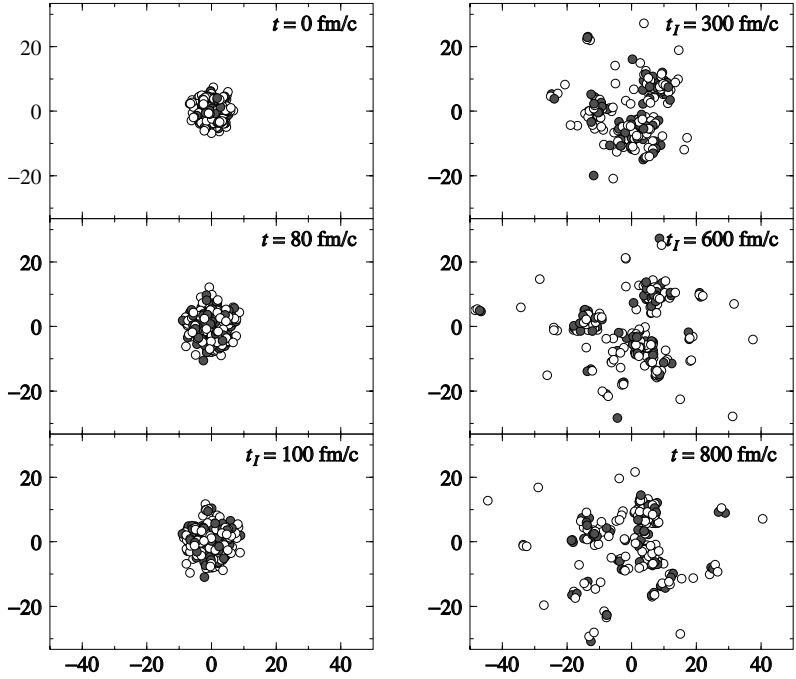}
\caption{Time evolution of $^{238}U$ at 6.8 MeV/nucleon excitation energy. 
At time $t=100$ fm/c the system enters imaginary time and exits at $t=600$ fm/c.
The real- and imaginary-times are indicated by $t$ and $t_I$.
}
\end{figure}
  Imagine that in some way we have been able to prepare the 
system at density $\rho$ and
temperature $T$.  Because of the compression and/or thermal pressure, the 
system will expand.  If
the excitation energy is too low the expansion will come to an halt and 
the system will shrink back. This
is some kind of monopole oscillation.   On the other hand if the 
excitation is very large it will
quickly expand and reach a region where the system is unstable and many 
fragments are formed, cf. Fig.~1.  It is
clear that in the expansion process the initial temperature will also 
decrease.  We could roughly
describe  this process with a collective coordinate $R(t)$, the radius of 
the system at time $t$ and its
conjugate coordinate.  Here we are simply assuming that the expansion is 
spherical.  These coordinates
 are somewhat the counterpart of the relative distance between fragments 
in the fission process. 
Similarly to the fission process we can imagine that connected to the 
collective variable $R(t)$ there
is a collective potential $V(R)$~\cite{fission}.  
When the excitation energy is too low it means that we are below
 the maximum of the potential.  That such a maximum exists comes, 
exactly as in the fission process,
from the short range nature of the nucleon-nucleon force  
 and the long range nature of Coulomb.  Thus, 
similarly to SF, we can imagine to
reach fragmentation by tunneling through the collective potential $V(R)$. 
 When this happens, fragments
will be formed at very low $T$ not reachable otherwise than through the 
tunneling effect.  The price
to pay as in all the subbarrier phenomena is the very low cross 
sections. 
Since the expansion of the system can be parametrized in one-dimensional
collective coordinate, we can, in principle, apply the imaginary-time 
method to its study similarly to ref.\cite{fission}.

Here we combine the imaginary time prescription with Quantum 
Molecular Dynamics \cite{qmd} model
to simulate fragmentation of finite system with relatively low 
excitation.
We simulate the expansion of $^{230}\rm U$ system.
First we prepare the ground state of a nucleus and then compress 
uniformly to
give an excitation energy $E^*$ from 5 to 8 MeV/nucleon.
Due to the fluctuations between events caused by the different initial 
configuration
of nucleus, the potential energy during the expansion is different for 
different events.
Therefore the tunneling fragmentation occurs in some events
where the potential energy is eventually high,
while there is no tunneling for events with lower potential energy.
The number of events with tunneling fragmentation is larger
for lower excitation energy.
In Fig.~3 we display snapshots of a typical tunneling event.  The 
collective coordinate $P(t)$
becomes zero at $t=100$ fm/c.  At this stage we turn to imaginary times 
 and the
tunneling begins.  Notice that the system indeed expands and its shape 
can be rather well approximated
to a sphere at the beginning.  But already at 300 fm/c (in imaginary 
time) due to the molecular dynamics
nature of the simulation, the spherical approximation is quite bad.  
This shortcoming of
our approach should be kept in mind because the calculated action will 
be quite unrealistic due to the
approximation used.  This is similar to the use of one collective 
coordinate (the relative distance
 between centers) in SF process.  In that case many calculated features 
are quantitatively wrong but
qualitatively acceptable.  Since our Quantum Drops is a proposed novel 
mechanism we can only give
qualitative features and the model assumption must be refined when 
experimental data will start to
be available~\cite{qdrop}.

\section{Conclusions}                  

Some theoretical indications for the behavior of a finite excited system are clear.  Both CMD and
BNV calculations of the MLE show that the system cannot hold more than
a certain temperature.  This seems intuitively reasonable because it
is quite clear that the finite nuclear system is not able to stay bound
when an excitation energy larger than 1.5$\sim$2 times the BE is given to it.
It seems that at such large excitation energies, the systems quickly
disassembles and consequently
the biggest fragment left will have a lower excitation
energy. The concept of a limiting temperature that the system
can sustain seems reasonable, and such a limiting temperature is strictly
related to the maximum value of the Lyapunov exponent
.
We propose also to search for fragmentation at very low excitation energies where quantum effects 
play a dominant role.  These rare events should give valuable informations about the instability region.

  %%%%%%%%%%%%%%%%%%%%%%%%%%%%%%%%%%%%%%%%%%%%%%%%%%%%%%%%%%%%%


\section*{References}
 \begin{thebibliography}{99}

\bibitem{ref1}  V.~Latora, M.~Belkacem and A.~Bonasera,
Phys. Rev. Lett. {\bf 73}%
,1765 (1994);
M.~Belkacem, V.~Latora and A.~Bonasera, Phys. Rev. C
52 (1995) 271;
P.~Finocchiaro, M.~Belkacem, T.~Kubo, V.~Latora and A.~Bonasera,
Nucl. Phys. {\bf A600}, 236 (1996).

\bibitem{ref2}  L.~Landau and E.~Lifshits, {\it Statistical Physics},
Pergamon, New York, 1980; \\ K.Huang, {\it Statistical Mechanics}, J.Wiley ,
New York, 1987, 2nd ed.

\bibitem{ref3} A.L.Goodman, J.~I.~Kapusta and A.~Z.~Mekjian, Phys. Rev. C
{\bf 30}, 851 (1984).

\bibitem{schu}A.Bonasera, Phys. World (Feb. 1999) 20.  A.~Bonasera and J.~Schulte, contr. to the Proceedings
  {\it Similarities and
Differences between Atomic Nuclei and Clusters},
 {\it et al.} Abe et al. editors, AIP (1998).

\bibitem{eos}  M.~L.~Gilkes et al.~Phys.~Rev.~Lett. {\bf 73}, 1590 (1994).

\bibitem{fis} M.~E.~Fisher, 
Proc. International School of Physics, Enrico Fermi Course LI, Critical
Phenomena, ed. M.~S.~Green (Academic, New York, 1971);
{1967}{255}.

\bibitem{cam86}  X.~Campi, J.~of Phys. {\bf A19}, 917 (1986); Phys. Lett.
{\bf B208}, 351 (1988).

%\bibitem{dor}C.~Dorso and A.~Strachan, Phys.Rev.{\bf B54}, 236 (1996).

\bibitem{mast}  P.~F.~Mastinu et al. Phys.Rev.Lett. {\bf 76}, 2646 (1996),
and references therein.


\bibitem{belk} M.~Belkacem et al. Phys.Rev. {\bf C54}, 2435 (1996).

\bibitem{dzp} V.~Latora, A.~Del~Zoppo and A.~Bonasera,
Nucl. Phys. {\bf A572}, 477 (1994).

\bibitem{rnc}A.~Bonasera, M.~Bruno, C.~Dorso and P.~F.~Mastinu, Riv.
Nuovo Cimento {\bf 23}, 1-101 (2000).

%\bibitem{dago} M.~D'Agostino et al. in preparation.

\bibitem{pp2}  A.~Bonasera, V.~Latora and A.~Rapisarda, Phys. Rev. Lett.
{\bf 75}, 3434 (1995).

\bibitem{burgio} G.~F.~Burgio and A.~Bonasera, in preparation.

\bibitem{hagel} K.~Hagel et al., Phys. Rev. {\bf 62}, 34607 (2000). 

\bibitem{fission} A.~Bonasera and A.~Iwamoto Phys. Rev. Lett. 
{\bf 78}, 187 (1997). 

\bibitem{qmd} T.~Maruyama et al., Phys. Rev. {\bf 57}, 655 (1998). 

\bibitem{qdrop} T.~Maruyama, S.~Chiba and A.~Bonasera, in preparation.
\end{thebibliography}
\end{document}